\documentclass[a4,german]{article}
\usepackage{cite} 
\usepackage{epsfig}
\usepackage{caption}
\usepackage{psfig}
\usepackage{amssymb,bbm}

\newcommand{\Rt}{\mb{Re}}

\newcommand{\ket}[1]{\, | #1 \rangle}

\newcommand{\Hnorm}{\mb{H}}

\newcommand{\mb}[1]{{\rm #1}}

\newcommand{\om}{\omega _0}
\newcommand{\Pbt}{P_{\rm bound}(t)}

\newcommand{\Pt}{P_{\rm Ion}(t)}
\newcommand{\he}{\hbar_{\rm eff}}

\newcommand{\Ge}{\Gamma_{\epsilon}}
\newcommand{\we}{w_{\epsilon}}

\begin{document} 


\section*{Anderson Localization in Atoms}
\subsection*{Sandro Wimberger$^1$ and Andreas Buchleitner$^2$}
\subsubsection*{{\it Max-Planck-Institut f\"ur Physik komplexer Systeme,
N\"othnitzer Str. 38, D-01187 Dresden\\
http://www.mpipks-dresden.mpg.de/mpi-doc/buchleitnergruppe/start.html\\
$^1$saw@mpipks-dresden.mpg.de, Tel.: 0351/871-1204, Fax: 0351/871-2299\\
$^2$abu@mpipks-dresden.mpg.de, Tel.: 0351/871-2211, Fax: 0351/871-2299}}


We investigate the celebrated analogy \cite{FGP82} between Anderson
localization phenomena in condensed matter physics \cite{KM93} and related
effects in the ionization process of driven Rydberg atoms, and put them on a
quantitative footing. \\
The atomic system we consider for a comprehensive numerical analysis of its
ionization properties is a highly excited hydrogen atom 
exposed to a monochromatic   
electromagnetic field of linear polarization. The external field is purely
classical, we ignore relativistic corrections, and assume an infinite proton
mass. Dropping the -- for our purposes negligible -- ponderomotive energy shift
\cite{Fai87}, and restricting to the dipole approximation, the Hamiltonian (in
atomic units, and applying the velocity gauge \cite{Fai87}) reads as
\begin{equation}
 \Hnorm   =  \frac{1}{2}{\vec{p}}^{\,2} -\frac{1}{r} -  \frac{Fp_z}{\omega}
\sin(\omega t). 
\label{hamilton2}
\end{equation}
Our method of a
combined application of the Floquet theorem and of complex scaling
\cite{BDG95} allows an exact description of the quantum dynamics of the
hydrogen atom, and fully includes the Coulomb singularity, 
as well as the field-induced coupling to the atomic continuum. 
Our only restriction here is a confinement of the electron 
to a one-dimensional configuration space along the polarization axis of the 
external field. 
The approximation of reduced dimensionality is reasonable for
quasi-one-dimensional Rydberg states for which the electronic motion is
essentially confined along the polarization axis \cite{BD97}. \\
The coupling to the atomic continuum
induced by the external field forces all bound states to turn into
resonances. These resonances 
are eigensolutions of the associated scattering problem, which satisfies
purely outgoing (Siegert-)boundary conditions \cite{Sie39}, and 
they are characterized by quasi-energies
$\epsilon$ with a resonance width $\Ge$. In our approach, these quasi-energies
are obtained by the method of 
complex scaling, and are given by the complex eigenvalues
$\epsilon=\Rt(\epsilon)-i\Ge/2$ of the scaled Floquet Hamiltonian.
The ionization probability of the initial -- field free -- Rydberg state $\ket
{n_0}$ with principal quantum number $n_0$ is completely defined by the
ionization widths (or rates) $\Ge$ of the individual Floquet eigenstates of
the atom in the field. Assuming rapid switching of the microwave field,
the weights $\we$ of the individual Floquet states are simply given by their
time-averaged overlaps with $\ket {n_0}$. Consequently, the
ionization probability reads as \cite{BDG95}
\begin{eqnarray}
\Pt  & =   & 1 -  \Pbt 
\nonumber  \\
    & =   & 1 -  \sum_{\epsilon} e^{- \Ge t} w_{\epsilon},
\label{pionisation}
\end{eqnarray} 
where $t$ is the total interaction time of the atom with the driving field.
The sum in (\ref{pionisation}) runs over the entire Floquet spectrum. \\
In the same way as the electronic transport through a disordered solid is
described by its transmission properties or the conductance \cite{KM93},
respectively, we define an atomic conductance $g$ \cite{BGZ98} 
to allow for a statistical characterization of the energy transport in
periodically driven atoms. 
$g$ is given by the weighted sum of the ionization rates $\Ge$, which
represent the atomic analogue of the transmission rates through a disordered
solid-state sample. From (\ref{pionisation}), it follows that  
\begin{equation}
g \equiv \frac{1}{\Delta} \sum_{\epsilon}\Gamma_{\epsilon}w_{\epsilon} = 
\left. \frac{1}{\Delta} \frac{d}{dt}\right| _{t \simeq 0} \Pt, 
\label{atomconduct1}
\end{equation}
with $\Delta$ being the mean level spacing of the Floquet spectrum. 
An estimate for $\Delta$ is obtained \cite{BGZ98} by
\begin{equation}
\Delta \equiv \frac{\omega}{N}, 
\label{leitniveauabstand}
\end{equation}
where $N \equiv 1/(2n_0^2 \omega )$ 
counts the number of photons the atom must absorb in order to ionize.

\begin{figure}
\centerline{\psfig{figure=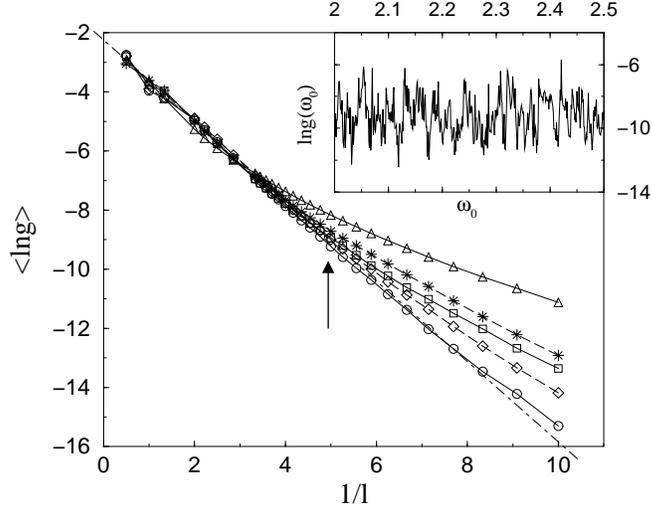,width=7cm,angle=270}}
\caption{Averaged values of the natural logarithm of the atomic conductance 
$g$ vs. the inverse localization ratio $1/\ell$ (initial states $n_0=40$
(triangles), 60 (stars), 70 (squares), 90 (diamonds), 100 (circles),
$\ell=0.1-2$). Each data point is obtained by averaging over an ensemble of 500
different samples (with the frequency measured in units of the classical,
undisturbed Kepler frequency $\om \equiv \omega \times n_0^3 
\in [2.0;2.5]$). For the initial principal quantum 
number $n_0=100$ we observe an almost linear dependence along the inserted
straight line (dashed-dotted), a result which is also obtained for disordered
solid-state models in the localized regime \cite{CGM97}. There is a clear
trend to approach the linear 
behaviour -- which one expects from the theory of dynamical localization -- as
$n_0$ increases from 40 to 100. The inset shows the huge fluctuations of $g$
-- on a logarithmic scale! -- as a function of the external (scaled) 
frequency, for $\ell =0.2$ and $n_0=100$ (corresponding to the data point
marked by the arrow in the main figure).} 
\label{fig1}
\end{figure}

\begin{figure}
\centerline{\psfig{figure=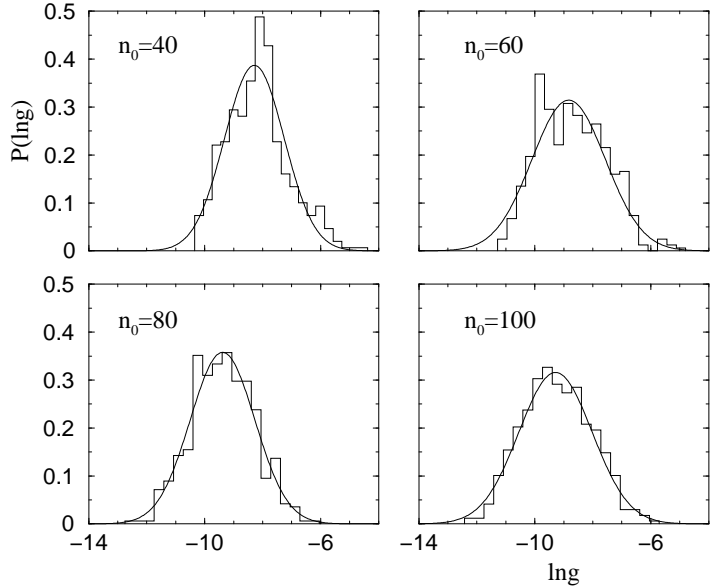,width=8cm,angle=270}}
\caption{Direct comparison of the statistical distributions of $\ln g$
(histograms) and the corresponding Gaussian fits (thick lines), 
for the initial states $n_0=40, 60, 80, 100$
(localization ratio $\ell = 0.2$). As the principal quantum number
$n_0$ increases at constant $\ell$, 
a manifest improvement of the approximation by the
fits, and a trend to shift to lower mean values is visible. Similar results
are obtained for the entire localized region $\ell < 0.5$.}
\label{fig2}
\end{figure}

Despite the fully deterministic dynamics which are generated by
(\ref{hamilton2}), the widths $\Ge$ and the overlaps $\we$ are very sensitive
to slight changes of the field parameters. This behaviour forces the
ionization probability and the conductance, respectively, to strongly fluctuate
as we vary, for instance, the microwave frequency. The observed huge
fluctuations (cf. inset in fig. \ref{fig1}) are the analogue of conductance
fluctuations known from condensed matter physics \cite{KM93}, where disorder
induced interference effects lead to an exponential 
localization of the electronic
wave function along the lattice, and therefore to a strong reduction of the
transmission through the solid. In both, the atomic and the solid-state
problem, a statistical description is needed for characterizing this complex
transport behaviour. Extreme fluctuations typically occur in a regime where
the overall -- atomic or mesoscopic -- conductance is exponentially
small. Then, the theory of dynamical localization \cite{CGS88}, 
which was developed in close analogy to the concept
of Anderson localization in condensed matter physics, predicts an exponential
decrease of the bound-state population distribution towards the atomic
continuum threshold.
This decay is completely characterized by the ratio of the localization
length $\xi$ (which measures the width of the bound-state distribution) 
to $N$. We quantitatively
check this prediction of dynamical localization by plotting the logarithmic
atomic conductance $\ln g$ vs. the inverse ratio $\ell ^{-1} \equiv N/\xi$.
For very high initial Rydberg sates, with principal quantum numbers close to 
100, we observe an exponential decrease of the atomic conductance
(cf. fig. \ref{fig1}). However, for smaller $n_0$, our results clearly show
systematic deviations from the exponential behaviour. The same deviations are
observable by looking at the statistical properties of the atomic conductance
\cite{Wim00}. 
Keeping the ratio $\ell$ fixed and changing the frequency of the externally
applied field in 500 equidistant, small steps, we obtain an ensemble of
500 different ``samples'' for each value of $n_0$ and $\ell$. Fig. \ref{fig2}
shows the log-normal distributions of the atomic conductance, which is in good
agreement with the predictions of solid-state models \cite{PZIS90}. Again, the
deviation of the distributions are the larger from the expected log-normal
behaviour, the smaller we have chosen the initial principal
quantum number $n_0$. 

The crucial insight gained from our study of the statistical properties of the
atomic conductance $g$ is that the initial Rydberg state defines an 
additional parameter $n_0 \sim \he
^{-1}$ \cite{KL95} (as compared to the scenario in Anderson-localized solids)
which non-trivially affects the statistical behaviour of  
the atomic transport process. Furthermore, the observed fluctuations prove to
be quite robust with respect to slight uncertainties in the applied field
strengths (the main problem in laboratory experiments \cite{SAKW94}), calling
for a direct experimental verifications of our numerical results. 


\end{document}